\documentclass[english,9 pt,a4paper]{article}
\usepackage{amsmath,amssymb}
\usepackage[square, comma, sort&compress,numbers]{natbib}
\usepackage{babel}
\usepackage{amsthm}
\usepackage{graphicx}
\usepackage[hmargin=2cm,vmargin={2.5cm,2cm}]{geometry}
\usepackage[final]{showkeys}
\usepackage{epstopdf}
\numberwithin{equation}{section}
\allowdisplaybreaks
\newtheorem{example}{Example}
\newtheorem{Theorem}{Theorem}[section]
\newtheorem{Remark}{Remark}[section]

\newtheorem{Lemma}{Lemma}[section]

\title{Analytic solution of nonlinear fractional Burgers-type equation by invariant subspace method}

\begin{document}

\author{Pietro Artale Harris\footnote{Email address: pietro.artaleharris@sbai.uniroma1.it}, Roberto Garra\footnote{Email address: roberto.garra@sbai.uniroma1.it}, \\
    Dipartimento di Scienze di Base e Applicate per l'Ingegneria, \\
    ``Sapienza'' Universit\`a di Roma\\
    Via A. Scarpa 16, 00161 Rome, Italy.\\
}
\maketitle

\begin{abstract}
In this paper we study the analytic solutions of  Burgers
type nonlinear fractional equations by means of the Invariant Subspace Method. We first study a class of nonlinear
equations directly related to the time-fractional Burgers equation.
Some generalizations linked to the forced time-fractional Burgers equations and variable-coefficient
diffusion are also considered. Finally we study a Burgers-type equation involving both space and time-fractional
derivatives.
\smallskip

\end{abstract}

\section{Introduction}

The analysis and applications of fractional differential equations
are object of an increasing interest in many different fields from
the viscoelasticity to the fluid mechanics (see for example
\cite{Podl}). In this framework much work has to be done to find
analytical methods to solve nonlinear fractional differential
equations. In literature there are different approximate and
semi-analytical methods to treat these equations, but in few cases
it is possible to find exact solutions. A classical nonlinear
equation is the Burgers equation, firstly introduced by J.M Burgers
in 1948 in the framework of the theory of turbulence \cite{bur}.
It is well known that this equation is linearizable to the heat
equation by using the Cole-Hopf transform. The time-fractional
Burgers equation was firstly treated by Momani in \cite{Momani} by
the Adomian decomposition method. More recently, other authors
studied this equation by using novel semi-analytical methods such as
the homotopy perturbation method \cite{Hom} and the Variational
Iteration Method \cite{VIM}. \\
The aim of this paper is to show an analytic method to find exact
solutions to nonlinear fractional Burgers-type equation, involving
time-derivatives in the Caputo sense.\\
Instead of directly studying the time-fractional Burgers equation, we study the following time-fractional nonlinear diffusion equation
\begin{equation}
\partial_t^\alpha u+\frac{1}{2}(\partial_x u)^2-k\partial_{xx}u=0.
\label{NL}
\end{equation}
In fact, the relation between these two equations is quite immediate: it suffices to differentiate equation \eqref{NL} with respect to $x$ to obtain
\begin{equation}
\partial_t^\alpha\partial_x u+\partial_x u\partial_{xx}u-k\partial_{xxx}u=0,
\end{equation}
then, by setting
\begin{equation}
f(x,t):=\partial_x u(x,t),
\label{relaz}
\end{equation}
we get the time-fractional Burgers equation:
\begin{equation}\label{ad}
\partial_t^\alpha f+f\partial_x f-k\partial_{xx}f=0.
\end{equation}
At this point we can find the exact solution to the time-fractional Burgers equation by \eqref{relaz}.
\\\indent In particular, we find exact
solution to equation \eqref{NL} by using
the Invariant Subspace Method, firstly introduced by Galaktionov
\cite{Gala} and recently used in the framework of nonlinear fractional equations
 by Gazizov and Kasatkin \cite{Gazizov}. A similar invariant analysis of the time-fractional
  Burgers equation was recently developed by Sahadevan and Bakkyaraj in \cite{india}; in this paper the authors have shown, by means of similarity transformation, the relation between
  time-fractional Burgers equation and nonlinear ordinary differential equations involving Erderly-Kober
  operator.\\
  We also study with the same method some generalizations of the time-fractional Burgers equation. \\
  A first generalization is given by considering in the fractional equation an elastic forcing term. \\
  Then, we study the fractional equation with time-dependent diffusion coefficient, i.e.
 \begin{equation}
\partial_t^\alpha u+\frac{1}{2}(\partial_x u)^2-k(t)\partial_{xx}u=0.
\end{equation}
For the sake of completness, we show that the Invariant Subspace Method provides exact solutions also
for a Burgers-type equation involving both space and time-fractional Caputo derivatives.

\bigskip

\section{Preliminaries}
In this section we give some basic notions about fractional calculus, and then we discuss the invariant subspace method as introduced by Galaktionov \cite{Gala}.

\subsection{Notations about Fractional Calculus}
Here we recall definition and basic results about fractional calculus, for more details we refer to \cite{Podl}.

\bigskip

Let $\gamma\in\mathbb{R}^{+}$, the Riemann-Liouville fractional integral is defined by
\begin{equation}
J^{\gamma}_t f(t) =
\frac{1}{\Gamma(\gamma)}\int_0^{t}(t-\tau)^{\gamma-1}f(\tau) d\tau,
\label{riemann-l}
\end{equation}
where
$$\Gamma(\gamma)= \int_0^{+\infty}x^{\gamma-1}e^{-x}dx,$$
is the Euler Gamma function.\\
Note that, by definition, $J^0_t f(t)= f(t)$. Moreover it satisfies the semigroup property, i.e. $J_t^{\alpha}J_t^{\beta} f(t)= J^{\alpha+\beta}f(t)$.\\
There are different definitions of fractional derivative (see e.g. \cite{Podl}). In this paper we used the fractional derivatives in the sense of Caputo.
Hereafter we denote by $AC^n([0,t])$, $n\in\mathbb{N}$, the class of functions $f(x)$ which are
continuously differentiable in $[0,t]$ up to order $n-1$ and with $f^{(n-1)}(x)\in AC([0,t])$.
We recall the following Theorem \cite[pag.92-93]{Kilbas}
\begin{Theorem}
Let $m-1 < \gamma< m$, with $m\in\mathbb{N} $. If $f(t)\in AC^n([0,t])$, then the Caputo
fractional derivative exists almost everywhere on $[0,t]$ and it is represented in the form
\begin{equation}
D_t^{\gamma}f(t)=   J^{m-\gamma}_t D_t^m f(t)=
\frac{1}{\Gamma(m-\gamma)}\int_0^{t}(t-\tau)^{m-\gamma-1}\frac{d^m}{dt^m}f (\tau) \, \mathrm d\tau, \;\gamma \ne m.
\end{equation}
\end{Theorem}
It is clear by definition that the fractional derivative is a pseudodifferential operator given by the convolution
of the ordinary derivative of the function with a power law kernel. So the reason why fractional derivatives introduce a memory formalism becomes evident.

It is simple to prove the following properties of fractional derivatives and integrals (see e.g. \cite{Podl}) that will
be used in the analysis:
\begin{align}
&D_t^{\gamma} J_t^{\gamma} f(t)= f(t), \quad \gamma> 0,\\
&J_t^{\gamma} D_t^{\gamma} f(t)= f(t)-\sum_{k=0}^{m-1}f^{(k)}(0)\frac{t^k}{k!}, \qquad \gamma>0, \: t>0,\\
&J_t^{\gamma} t^{\delta}= \frac{\Gamma(\delta+1)}{\Gamma(\delta+\gamma+1)}t^{\delta+\gamma} \qquad \gamma>0, \: \delta>-1, \: t>0,\\
&D_t^{\gamma} t^{\delta}= \frac{\Gamma(\delta+1)}{\Gamma(\delta-\gamma+1)}t^{\delta-\gamma} \qquad \gamma>0, \: \delta>-1, \: t>0.
\end{align}

\subsection{Invariant subspace method}

The invariant subspace method, as introduced by Galaktionov \cite{Gala}, allows to solve exactly nonlinear equations by separating variables.\\
Only recently Gazizov and Kasatkin \cite{Gazizov} suggested its application to nonlinear fractional equations. \\
We recall the main idea of this method: consider a scalar evolution
equation
\begin{equation}\label{pro}
\frac{\partial u}{\partial t}= F[u],
\end{equation}
where $u=u(x,t)$ and $F[u]$ is a nonlinear differential operator. \\
Given $n$ linearly independent functions
$$f_1(x), f_2(x),....,f_n(x),$$
we call $W_n$, the $n$-dimensional linear space
$$W_n=\langle f_1(x), ...., f_n(x)\rangle.$$
This space is called invariant under the given operator $F[u]$, if
$F[u]\in W_n$ for any $u\in W_n$. This means that there exist $n$
functions $\Phi_1, \Phi_2,..., \Phi_n$ such that
$$F[C_1f_1(x)+...+C_n f_n(x)]= \Phi_1(C_1,...,C_n)f_1(x)+...+\Phi_n(C_1,...,C_n)f_n(x),$$
where $C_1, C_2, ..., C_n$ are arbitrary constants. \\
Once the set of functions $f_i(x)$ that form the invariant subspace
is given, we search an exact solution of \eqref{pro} in the
invariant subspace in the form
\begin{equation}
u(x,t)=\sum_{i=1}^n u_i(t)f_i(x).
\end{equation}
where $f_i(x)\in W_n$. In this way, we arrive to a system of ODEs.
In many cases this is a problem -simpler than the original one- that allows
to find exact solutions by just separating variables \cite{Gala}.
The same method can be applied to nonlinear time-fractional equations (see \cite{Gazizov}). In this case, when the invariant subspace has been found,
the nonlinear equation is reduced to a system of time-fractional ODEs.\\

\bigskip

\section{The time-fractional Burgers equation}
In this section we study the solution and the meaning of the time-fractional Burgers equation by means of relation \eqref{relaz} with the non linear fractional diffusion equation \eqref{NL}.
\subsection{Analytic solution of the nonlinear diffusion equation by Invariant Subspace Method}

Let us write again the time-fractional nonlinear diffusion equation (equation \eqref{NL})
\begin{equation}
\partial_t^\alpha u+ \frac{1}{2}\left(\partial_x u\right)^2 -k
\partial_{xx}u=0.
\label{Burg1}
\end{equation}
Here $u=u(x,t)$ is a "good" function, that is derivable in the sense
of Caputo (see Theorem 2.1), $x \in I\subseteq\mathbb{R},t\geq
0,\;\alpha\in(0,1)/\left\lbrace\frac{1}{2}\right\rbrace$ and
$\partial_t^\alpha$ is the time-fractional derivative in the Caputo
sense. The reason why we must exclude $\alpha =1/2$ will be clear in the following.\\
It is possible to find an exact solution to \eqref{Burg1} by using the Invariant subspace method.
In \eqref{Burg1} we have
$$F[u]=-\frac{1}{2}\left(\partial_x u\right)^2+ k \partial_{xx}u.$$
In the following, we take $k=1$ for simplicity.
We consider the subspace $W_3=\langle 1, x, x^2\rangle$. We have that
$W_3$ is preserved by $F[u]$, since
\begin{equation}
F(C_1+ C_2 x+C_3x^2)= -\frac{1}{2}C_2^2- 2C_3^2x^2-2C_2C_3x+2C_3 \in W_3,
\end{equation}
where the dimension of $W_3$ is three. All this allows us to find an
exact solution as
\begin{equation}\label{pis}
u(x,t)= a(t)+b(t)x+c(t)x^2.
\end{equation}
By substitution of \eqref{pis} in \eqref{Burg1}, we therefore obtain
\begin{equation}
\partial_t^\alpha a(t)+x\partial_t^\alpha b(t)+x^2\partial_t^\alpha c(t)= -\frac{1}{2}b(t)^2-2x^2c(t)^2-2xb(t)c(t)+2 c(t).
\label{Burg2}
\end{equation}
Hence we have a simple system of ODEs:
\begin{eqnarray}
\begin{cases}
\frac{d^{\alpha}a(t)}{dt^{\alpha}}=2c(t)-\frac{1}{2}b(t)^2,\\ &\\
\frac{d^{\alpha}b(t)}{dt^{\alpha}}=-2b(t)c(t),\\&\\
\frac{d^{\alpha}c(t)}{dt^{\alpha}}=-2c(t)^2.
\end{cases}
\end{eqnarray}

By using the notions as given in section 2, we can find that a
solution of the third equation is simply given by
\begin{equation}\label{ri1}
c(t)= -\frac{1}{2}\frac{\Gamma(1-\alpha)}{\Gamma(1-2\alpha)}t^{-\alpha}.
\end{equation}
By substituting in the second equation we find that
$$\frac{d^{\alpha}b(t)}{dt^{\alpha}}=b(t)\frac{\Gamma(1-\alpha)}{\Gamma(1-2\alpha)}\frac{1}{t^{\alpha}},$$
whose solution is simply given by
\begin{equation}\label{ri2}
b(t)=t^{-\alpha}.
\end{equation}
Substituting in the first equation, we get
\begin{equation}\nonumber
\partial_t^\alpha a(t)=-\frac{1}{2}t^{-2\alpha}-
\frac{\Gamma(1-\alpha)}{\Gamma(1-2\alpha)}t^{-\alpha},
\end{equation}
 and thus,
\begin{equation}\label{ri3}
a(t)=-\frac{1}{2}\frac{\Gamma(1-2\alpha)}{\Gamma(1-\alpha)}t^{-\alpha}-\left(\frac{\Gamma(1-\alpha)}{\Gamma(1-2\alpha)}\right)^2.
\end{equation}
We finally find a complete exact solution
\begin{equation}\label{solu}
u(x,t)=t^{-\alpha}\left(-\frac{1}{2}\frac{\Gamma(1-2\alpha)}{\Gamma(1-\alpha)}+x-\frac{1}{2}\frac{\Gamma(1-\alpha)}{\Gamma(1-2\alpha)}x^2\right)-\left(\frac{\Gamma(1-\alpha)}{\Gamma(1-2\alpha)}\right)^2.
\end{equation}
It is now clear that we must consider $\alpha \neq 1/2$ to avoid
the divergence due to the Gamma function singularity in zero.
Moreover, we observe that the sign of the solution depends on the
range of variability of the fractional parameter $\alpha$ appearing
in the Gamma functions. Indeed, the sign is positive if $\alpha>1/2$
and
negative if $\alpha <1/2$.\\
This analysis allows to find the autosimilarity solution of the
time-fractional nonlinear diffusion equation. Starting from this
analysis we can find also a solution of the time-fractional Burgers
equation in a similar way. Indeed we can find an exact solution to
\eqref{ad}, by using \eqref{solu}. We notice that the
corresponding solution is the inviscid selfsimilar solution of the
time-fractional Burgers equation, previously discussed for example
by Gazizov and Kasatkin in \cite{Gazizov}.\\
In order to understand the meaning of this particular exact
solution, we observe that it is strictly related to the rarefaction
wave solution of the Burgers equation. This suggests a similar
interpretation of this solution as a fractional rarefaction wave
that solves a fractional Cauchy problem  with an initial step
condition (see Remark 4.1).

\subsection{Fractional conservation of mass}

In this subsection, we discuss the meaning of \eqref{ad} with $k=0$.
In particular we show that the mass conservation is preserved also
by introducing a memory formalism in the Burgers equation by means
of fractional operators. First of all, we observe that
$$\partial_t^{\alpha}u(t)= J^{1-\alpha}_t \left(\partial_t u(t)\right),$$
and
$$\partial_t^{1-\alpha}J^{1-\alpha}_t = Id. $$
Then by applying the Caputo fractional derivative of order
$1-\alpha$ on both the sides of equation \eqref{ad} and taking
$k=0$, we have
\begin{equation}
\partial_t u(x,t)=-\partial_t^{1-\alpha}\partial_x
\left(\frac{u^2}{2}\right).
\end{equation}
Hereafter, we assume the boundary conditions $u(x\rightarrow \pm
\infty, t)=0$. Hence we have
\begin{align}
\partial_t\int_{-\infty}^{+\infty} u(x,t)dx &= - \partial_t^{1-\alpha}\int_{-\infty}^{+\infty} \partial_x
\left(\frac{u^2}{2}\right)dx\\
\nonumber &= -\partial_t^{1-\alpha}
\left[\frac{u^2}{2}\right]_{-\infty}^{+\infty} =0.
\end{align}
This proves that the mass conservation is preserved by the
introduction of a fractional approach in the Burgers equation.

\section{Generalizations}

\subsection{Time-fractional Burgers equation with elastic forcing}

We now consider the inviscid time-fractional Burgers equation, including in the master
equation a forcing term $V(x,t)$, i.e.
\begin{equation}
\partial_t^\alpha u+ u\partial_x u= V(x,t), \quad \alpha \in (0,1).
\label{Burg2}
\end{equation}
We study, as a special case of \eqref{Burg2},  $V(x,t)= K(t) x$, that is an elastic type forcing term, with
time-dependent elasticity. On the basis of the previous analysis,
we search the solution by
separating variable method, as follows
\begin{equation}
u(x,t)= a(t)+b(t)x.
\end{equation}
We have
\begin{equation}
\frac{d^{\alpha}a(t)}{dt^{\alpha}}+x\frac{d^{\alpha}b(t)}{dt^{\alpha}}+a(t)b(t)+xb^2(t)= K(t)x.
\end{equation}
Hence we have again a system of ODEs
\begin{align}\label{sist}
\begin{cases}
\frac{d^{\alpha}a(t)}{dt^{\alpha}}=-a(t)b(t),\\&\\
\frac{d^{\alpha}b(t)}{dt^{\alpha}}=-b^2(t)+K(t).
\end{cases}
\end{align}
The second equation is a fractional Riccati equation, whose solution clearly depends on
$K(t)$. There are many
recent studies about fractional Riccati equations (see for example
\cite{Riccati} and references therein), but in few cases it is
possible to find an analytic exact
solution.\\
In order to find an exact solution in a more general case, we will
use the following auxiliary
\begin{Lemma}
Let us consider a function
\begin{equation}
K(t)= h^2(t)+\frac{d^{\alpha} h(t)}{dt^{\alpha}},
\end{equation}
where $h(t)\in AC([0,t])$. Then, a solution of the Cauchy problem
\begin{equation}
\begin{cases}
\frac{d^{\alpha}b(t)}{dt^{\alpha}}= - b^2(t)+K(t),\\&\\
b(0)=h(0),
\end{cases}
\end{equation}
is given by
\begin{equation}
b(t)= h(t).
\end{equation}
\end{Lemma}
By means of this Lemma, we can find a wide class of solutions to the
system \eqref{sist}, once fixed $h(t)$.

\begin{example}
Let us consider the following Cauchy problem
\begin{equation}
\begin{cases}
\frac{d^{\alpha}b(t)}{dt^{\alpha}}= - b^2(t)+K(t),\\&\\
 b(0)=1,
\end{cases}
\end{equation}
with
$$K(t)= E_{\alpha}(t^{\alpha})(1+E_{\alpha}(t^{\alpha})),$$
where
$$E_{\alpha}(t)= \sum_{k=0}^{\infty}\frac{(t^{\alpha})^k}{\Gamma(k\alpha+1)},$$
is the Mittag-Leffler function. Then, its analytic solution is given
by
$$b(t)= E_{\alpha}(t^{\alpha}).$$

\end{example}

\subsection{Time-fractional Burgers equation with time-dependent diffusion coefficient}
In some recent papers (see for example \cite{poc} and references therein), a generalized Burgers equation was considered, by taking a time-dependent
diffusion coefficient in the master equation. In our framework this means to consider the equation
\begin{equation}
\partial_t^\alpha u+\frac{1}{2}(\partial_x u)^2-k(t)\partial_{xx}u=0,
\end{equation}
where $k(t)$ belongs to the set of nonvanishing smooth functions of $t$.
Up to now we will always consider $\alpha \neq 1/2$ for the same reason discussed above.
In this case the subspace $W_3=\langle 1, x, x^2\rangle$ is invariant and we can find the solution
as
$$u(x,t)= a(t)+b(t)x+c(t)x^2.$$
Then, by substitution we should study the following equation
$$\partial_t^\alpha a(t)+x\partial_t^\alpha b(t)+x^2\partial_t^\alpha c(t)= -\frac{1}{2}(b(t)^2- 2x^2c(t)^2-2 xb(t)c(t)+2 k(t) c(t).
$$
hence the problem is reduced to the analysis of a system of coupled nonlinear fractional
ordinary differential equations
\begin{align}
\begin{cases}
\frac{d^{\alpha}a(t)}{dt^{\alpha}}=2k(t)c(t)-\frac{1}{2}b(t)^2,\\&\\
\frac{d^{\alpha}b(t)}{dt^{\alpha}}=-2b(t)c(t),\\&\\
\frac{d^{\alpha}c(t)}{dt^{\alpha}}=-2c(t)^2.
\end{cases}
\end{align}
The solutions of this system of equations are achieved by
using essentially the same calculations of the previous sections.
Indeed solutions of the second and third equations are given by
\eqref{ri1} and \eqref{ri2} above. The solution of the first
equation can be found by fractional integration, i.e., assuming that
$a(0)=0$,
\begin{align}
a(t)&= -\frac{1}{2}J^{\alpha}b(t)^2+2J^{\alpha}k(t)c(t)\\
\nonumber &=
-\frac{\Gamma(1-2\alpha)}{2\Gamma(1-\alpha)}t^{-\alpha}-
\frac{\Gamma(1-\alpha)}{\Gamma(1-2\alpha)}J^{\alpha}\left(k(t)t^{-\alpha}\right).
\end{align}
Again, the sign of the solution depends on $\alpha$, as discussed
above. Clearly the solution depends by the particular choice of
$k(t)$. However it suffices that $k(t)$ belongs to $L_1([0,t])$, in
order to be integrable in the sense of Riemann-Liouville (see
\cite{Kilbas}) and the complete exact solution can be achieved.

\subsection{Burgers-type equations involving space and time fractional derivatives}

In recent papers space-fractional Burgers equations has been studied by different authors.
For example in \cite{Miskinis}, it was studied the following space-fractional Burgers equation
$$\partial_t u+\frac{1}{2}\partial_x^{\beta}(\partial_x^{1-\beta} u)^2-k\partial_{xx}u=0, \quad \beta \in [0,1].$$
This equation interpolates the classical Burgers equation for $\beta =1$ and the nonlinear diffusion
equation for $\beta = 0$. It was shown that it is linearizable by means
of a generalized Cole-Hopf transform. \\
In \cite{pippo}, the following equation was studied
\begin{equation}\label{pip}
\partial_t u+ \partial_x u\partial_x^{\beta}u =0,\quad \beta \in(0,1),
\end{equation}
in relation to the propagation of nonlinear thermo-elastic waves in
porous media (see also \cite{pippo1} and references therein for the
model equations).\\ Here we consider a Burgers-type equation involving
both the space and time-fractional Caputo derivatives, i.e.
\begin{equation}\label{mia}
\partial_t^{\alpha} u+ u\partial_x^{\beta}u =0,\quad \alpha \in(0,1)/\{1/2\}, \beta \in(0,1).
\end{equation}
It is simple to see that \eqref{pip} admits an invariant subspace given by
$$W_2=\langle 1, x^{\beta}\rangle$$
Then we can search an exact solution to \eqref{mia} in the form
$$u(x,t)=a(t)x^{\beta}+\mbox{const}.$$
By substitution in \eqref{mia}, we have
\begin{equation}
x^{\beta}\frac{d^{\alpha}}{dt^{\alpha}}a(t)+a^2(t)x^{\beta}\Gamma(\beta+1)=0,
\end{equation}
and finally
\begin{equation}
\frac{d^{\alpha}}{dt^{\alpha}}a(t)=-a^2(t)\Gamma(\beta+1).
\end{equation}
By simple calculations we conclude that
$$u(x,t)=-\frac{\Gamma(1-\alpha)}{\Gamma(\beta+1)\Gamma(1-2\alpha)}\frac{x^{\beta}}{t^{\alpha}}+\mbox{const.}$$

\section{Final remarks}
In this final section several remarks are discussed.
\begin{Remark}
It is trivial to note that, in the limit $\alpha\rightarrow 1$, $k \rightarrow 0$ in \eqref{ad}, the usual Burgers equation is recovered:
\begin{equation}
\partial_t u+u\partial_x u=0.
\label{burgusual}
\end{equation}
Consider now \eqref{burgusual} with the following initial condition:
\begin{align}
g(x)=
\begin{cases}
0\qquad\mbox{if}\qquad x<0\\
1\qquad\mbox{if}\qquad x>0.
\end{cases}
\label{condini}
\end{align}
A solution of \eqref{burgusual}, \eqref{condini} can be obtained by using the method of characteristics (see e.g. \cite{Evans}):

\begin{align}
u(x,t)=
\begin{cases}
1\qquad\mbox{if}\qquad x>t\\
\frac{x}{t}\qquad\mbox{if}\qquad 0<x<t\\
0\qquad\mbox{if}\qquad x<0.
\end{cases}
\end{align}
This argument suggests to us that, assuming the same initial condition \eqref{condini} for equation \eqref{ad} with $k=0$, we have the same domain of validity for its solution simply by substituting $t$ with $t^\alpha$, i.e.
\begin{align}
u(x,t)=
\begin{cases}
1\qquad\mbox{if}\qquad x>t^\alpha/C_\alpha,\\
C_{\alpha}\frac{x}{t^\alpha}\qquad\mbox{if}\qquad 0<x<t^{\alpha}/C_{\alpha},\\
0\qquad\mbox{if}\qquad x<0,
\end{cases}
\end{align}
where $C_{\alpha} = -\Gamma(1-\alpha)/\Gamma(1-2\alpha)$, with
$\alpha >1/2$.
\end{Remark}

\begin{Remark}
We observe that the concern of the existence and uniqueness of the solution
for the space and time fractional Burgers equation
$$\partial_t^{\alpha}u+u\partial_x u = \partial_x^{\beta}u, \quad x>0,\, t>0,$$
was studied by Rodrigues in \cite{Rod}
by using Banach fixed point theorem. It is simple to prove that similar arguments apply in our case.
\end{Remark}

\begin{Remark}
In this paper, we provide some applications of the invariant subspace method to the fractional Burgers-type equation.
We observe that this method can be applied also to more complicated fractional equations and the generator
of the invariant subspace can be classical special functions of the fractional calculus.\\
The study of these equations will be the matter of further investigations.
\end{Remark}

\end{document}